\documentclass[prl,reprint,twocolumn,showpacs]{revtex4-1}
\usepackage{amsmath,amssymb,amsfonts}
\usepackage{graphicx}
\usepackage{txfonts}
\usepackage[T1]{fontenc}
\usepackage{textcomp}
\usepackage{graphicx}
\usepackage{upgreek}
\usepackage{braket}
\usepackage{bbold}
\usepackage{bm}
\usepackage[mathscr]{euscript}
\usepackage[unicode]{hyperref}
\hypersetup{
    unicode=true,                                           
    plainpages=false,
    pdffitwindow=true,                                      
    pdftitle={Soliton Interferometry},                      
    pdfauthor={John Lloyd Helm},                            
    pdfsubject={},                                          
    pdfkeywords={Bose--Einstein, condensate, BEC, solitons},
    colorlinks=true,                                        
    linkcolor=blue,                                         
    citecolor=blue,                                         
    filecolor=blue,                                         
    urlcolor=blue                                           
}
\newcommand{\eqnrefp}[1]{{[Eq.~(\ref{#1})]}}

\newcommand{\eqnreft}[1]{{Eq.~(\ref{#1})}}
\newcommand{\eqnreftfull}[1]{{Equation~(\ref{#1})}}

\newcommand{\figreft}[2]{Fig.~\ref{#1}#2}

\newcommand{\figrefp}[2]{[Fig.~\ref{#1}#2]}

\newcommand{\mcsa}{(\textcolor[rgb]{0,0,0}{\bm{+}})}
\newcommand{\mcsb}{(\textcolor[rgb]{0,0,0}{\bm{\circ}})}
\newcommand{\mcsc}{(\textcolor[rgb]{0,0,0}{\bm{\triangle}})}
\newcommand{\mcsd}{(\textcolor[rgb]{0,0,0}{\bm{\square}})}
\newcommand{\mcse}{(\textcolor[rgb]{0,0,0}{\bm{\times}})}

\newcommand{\etal}{\emph{et al.}~}

\newcommand{\eye}{i}
\newcommand{\pT}{\phi_{\mathrm{T}}}
\newcommand{\pR}{\phi_{\mathrm{R}}}
\newcommand{\pC}{\phi_{\mathrm{C}}}

\newcommand{\pS}{\phi_{\mathrm{s}}}

\newcommand{\pa}{\partial}

\newcommand{\Sdelta}{\delta_{\mathrm{S}}}
\newcommand{\sech}{\mathrm{sech}}

\renewcommand{\cos}{\mathrm{cos}}

\renewcommand{\Im}{\operatorname{Im}}
\newcommand{\gd}{g_{\mathrm{1D}}}
\newcommand{\Tc}{T_{\mathrm{c}}}
\newcommand{\Ts}{T_{\mathrm{s}}}

\newcommand{\nb}{n_{\mathrm{b}}}

\begin{document}

\title{Sagnac Interferometry Using Bright Matter-Wave Solitons}
\author{J. L. Helm}
\affiliation{Joint Quantum Center (JQC) Durham--Newcastle, Department of Physics, Durham University, Durham DH1 3LE, United Kingdom}
\author{S. L. Cornish}
\affiliation{Joint Quantum Center (JQC) Durham--Newcastle, Department of Physics, Durham University, Durham DH1 3LE, United Kingdom}
\author{S. A. Gardiner}
\affiliation{Joint Quantum Center (JQC) Durham--Newcastle, Department of Physics, Durham University, Durham DH1 3LE, United Kingdom}
\date{\today}

\begin{abstract}
We use an effective one-dimensional Gross--Pitaevskii equation to study bright matter-wave solitons held in a tightly confining toroidal trapping potential, in a rotating frame of reference, as they are split and recombined on narrow barrier potentials. In particular, we present an analytical and numerical analysis of the phase evolution of the solitons and delimit a velocity regime in which soliton Sagnac interferometry is possible, taking account of the effect of quantum uncertainty.
\end{abstract}

\maketitle

A Bose--Einstein condensate (BEC) with attractive inter-atomic interactions can support soliton-like structures referred to as bright solitary matter-waves \cite{khaykovich_etal_science_2002, strecker_etal_nature_2002, cornish_etal_prl_2006, Marchant_etal_2013, Nguyen2014}.  These propagate without dispersion \cite{morgan_etal_pra_1997}, are robust to collisions with other bright solitary matter-waves and with slowly varying external potentials \cite{parker_etal_physicad_2008, billam_etal_pra_2011}, and have center-of-mass trajectories well-described by effective particle models \cite{martin_etal_prl_2007, martin_etal_pra_2008, poletti_etal_prl_2008}.  Such soliton-like properties are due to the mean-field description of an atomic BEC reducing to the nonlinear Schr\"{o}dinger equation in a homogeneous, quasi-one-dimensional (quasi-1D) limit, which for the case of attractive interactions supports the bright soliton solutions well-known in the context of nonlinear optics~\cite{zakharov_shabat_1972_russian, satsuma_yajima_1974, gordon_ol_1983, haus_wong_rmp_1996, Helczynski_ps_2000}. The quasi-1D limit is experimentally challenging for attractive condensates \cite{billam_etal_variational_2011}, but solitary wave dynamics remain highly soliton-like outside this limit \cite{cornish_etal_prl_2006, billam_etal_pra_2011}. 

A bright solitary wave colliding with a narrow potential barrier is a good candidate mechanism to create two mutually coherent localised condensates, much as a beam-splitter  splits the light of an optical interferometer. This has been extensively 
investigated in the quasi-1D, mean-field description of an atomic BEC~\cite{HELM_PRA_2012,kivshar_malomed_rmp_1989, ernst_brand_pra_2010,lee_brand_2006, cao_malomed_pla_1995, holmer_etal_cmp_2007, holmer_etal_jns_2007,POLO_etal_PRA_2013,Molmer_arxiv_2012,Minmar_thesis_2012,abdullaev_brazhnyi_2012}, and sufficiently fast collisions do lead to the desired beam-splitting effect \cite{holmer_etal_cmp_2007, holmer_etal_jns_2007}. Consequently, bright solitary matter-waves, with their dispersion-free propagation, present an intriguing candidate system for future interferometric devices \cite{HELM_PRA_2012,strecker_etal_nature_2002, cornish_etal_physicad_2009, weiss_castin_prl_2009, streltsov_etal_pra_2009, billam_etal_pra_2011, al_khawaja_stoof_njp_2011,martin_ruostekoski_njp_2010,mcdonald_prl_2014}. Previous work~\cite{martin_ruostekoski_njp_2010,POLO_etal_PRA_2013,HELM_PRA_2014} considered a Mach--Zehnder interferometer using a narrow potential barrier to split harmonically trapped solitary waves, based on the configuration of a recent experiment ~\cite{Nguyen2014}. These demonstrated one can also recombine solitary waves if they collide at the barrier; the collision dynamics are explained more fully in~\cite{HELM_PRA_2012}. In these collisions, the relative atomic populations within the two outgoing solitary waves are governed by the relative phase $\Delta$ between the incoming ones.  The mean-field nonlinearity can lead to the relative populations of the outgoing waves exhibiting greater sensitivity to small variations in the phase $\Delta$; however, simulations including quantum noise in the initial condition~\cite{HELM_PRA_2014} or via the truncated Wigner method~\cite{blakie_etal_ap_2008}, demonstrated that enhanced number fluctuations counteract this improvement~\cite{martin_ruostekoski_njp_2010}.

\begin{figure}[t]
  \centering
  \includegraphics[width=\columnwidth]{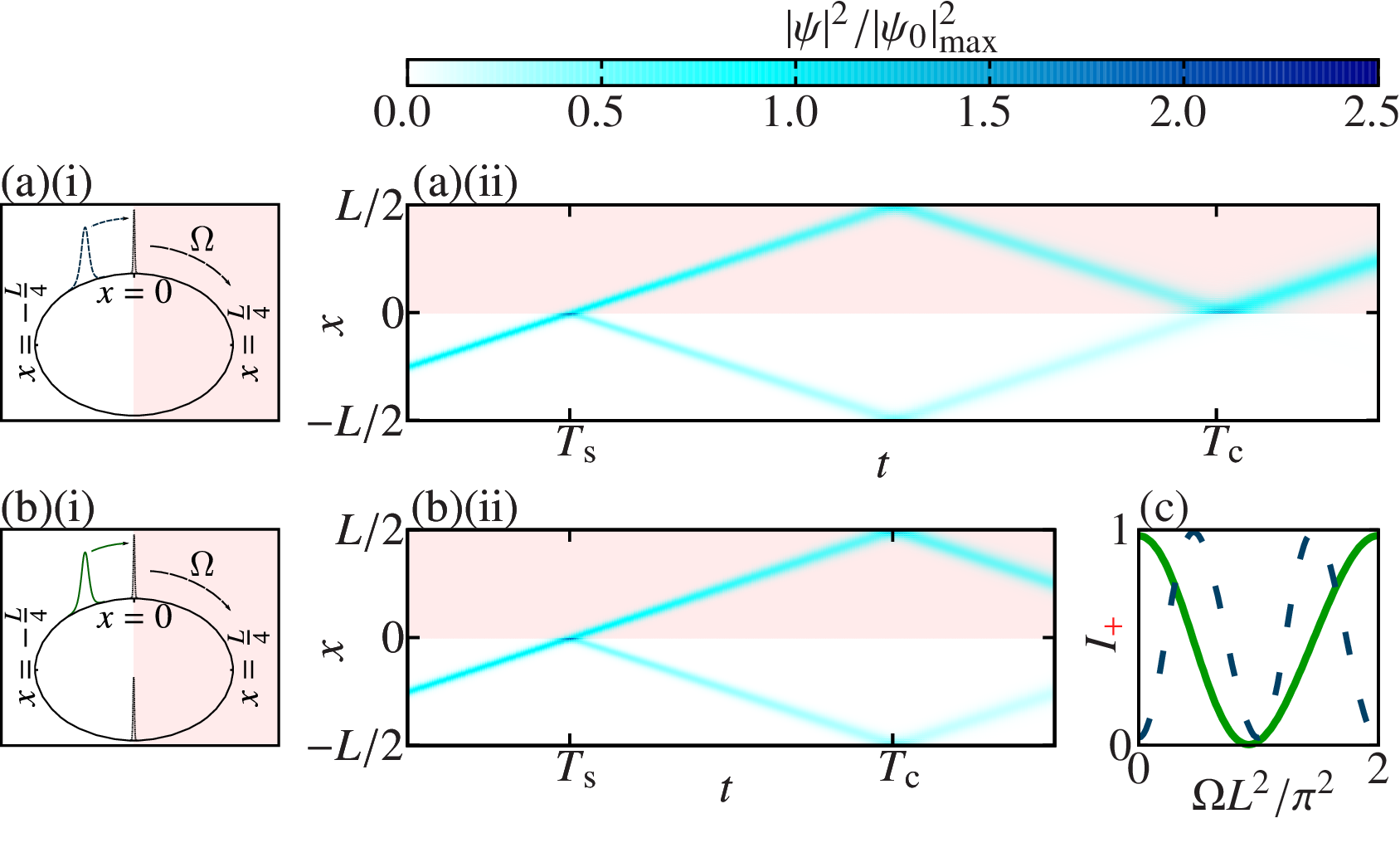}
  \caption[Diagram of Sagnac interferometry]{Stages of Sagnac interferometry.  An incoming soliton splits at time $T_{\mathrm{s}}$ on a barrier into two solitons of equal amplitude and opposite velocity.  After circumnavigating the ring trap, at time $T_{\mathrm{c}}$ the solitons recombine either at the same barrier (a), or a second barrier (b) antipodal to the first, illustrated in both cases with angular rotation $\Omega=1.875 \times 10^{-3}$, and ring circumference $L=40\pi$.  The resulting phase difference, incorporating the Sagnac phase due to the rotating reference frame, is read out via the population difference in the final output products within the positive (shaded) and negative domains.  (c) Final population in the positive domain $I_{+}$ as a function of $\Omega$, with $L=40\pi$ and initial soliton velocity $v=4$.  The sensitivity of the single barrier case (dashed line) is twice that of the double barrier case (solid line) because the interrogation time $T_{\mathrm{c}}-T_{\mathrm{s}}$ is doubled.}
\label{fig:overview} 
\end{figure}

We extend the framework of soliton interferometry to measurement of the Sagnac effect, first observed in an atom interferometer by Riehle \etal\cite{Riehle_etal_prl_1991}. In this experiment the observation manifested as a shift in the Ramsey fringes produced by passing an atomic beam of $^{40}$Ca through four travelling waves in a Ramsey geometry, producing an atomic beam interferometer.  What we present differs from the Riehle setup in two ways. Firstly, in~\cite{Riehle_etal_prl_1991} some phase information is transported optically. In our system
atom-light interactions serve only to coherently split the condensate; any resulting
phase dynamics are incidental. Secondly, our system results, not in an interference fringe shift, but a population shift between the positive and negative domains of the interferometer. The Sagnac effect is inferred from measurements of particle numbers~\cite{halkyard_etal_pra_2010} in the spatially distinct condensates on either side of the barrier, and not the structure of those condensates. 
(which are expected to remain soliton-like).
We consider an experimental configuration, contained entirely within a rotating frame, where there is a smooth ring-shaped trapping potential (implemented by, e.g., using a spatial light modulator \cite{moulder_etal_pra_2012}, time-averaging with acousto-optic deflectors \cite{henderson_etal_njp_2009}, or imaging an intensity mask \cite{corman_etal_prl_2014}) and narrow barriers realised with optical light sheets, focussed using high numerical aperture lenses.  Solitons, initially produced in an optical trap, can be adiabatically transferred into the ring, with the initial velocity set by moving the optical potential during the transfer \cite{rakonjac_etal_ol_2012}.  Key sources of error include: uncertainty in the value of the soliton velocity relative to the barrier strength and, in turn, the barrier transmission level \cite{HELM_PRA_2012}; initial particle number, which determines the ground-state energy and so sets the low-energy splitting threshold, close to which the system becomes sensitive to otherwise small fluctuations in the velocity \cite{martin_ruostekoski_njp_2010}; and measurement of final particle number.

Within the Gross--Pitaevskii equation (GPE) framework, we consider $N$ bosonic atoms of mass $m$ and scattering length $a_{s}$, in an effective 1D configuration due to a tightly confining (frequency $\omega_{r}$) harmonic trapping potential in the degrees of freedom perpendicular to the direction of free motion, implying an interaction strength of $\gd= 2\hbar\omega_{r}a_{s}$ per particle.  We use ``soliton units'' \cite{martin_etal_pra_2008} (equivalent to $\hbar = m = |\gd|N =1$), where position, time and energy are in units of $\hbar^{2}/m|\gd|N$, $\hbar^{3}/m\gd^{2}N^{2}$, and $m\gd^{2}N^{2}/\hbar^{2}$~\cite{Note1}. To describe a tightly confining toroidal trap geometry (or ring trap), we introduce periodic boundary conditions over the domain $-L/2<x\le L/2$, where $L$ is the dimensionless form of the circumference~\cite{HELM_PRA_2014}.  It is common to discuss Sagnac interferometry and ring systems in terms of an angle coordinate $\theta = 2\pi x/L$ \cite{halkyard_etal_pra_2010,kanamoto_etal_pra_2003}; we choose not to, making it easier to draw on earlier work on splitting solitons at narrow barriers \cite{HELM_PRA_2012,holmer_etal_cmp_2007,holmer_etal_jns_2007,HELM_PRA_2014}. Considering the dynamics within a frame rotating with dimensionless angular frequency $\Omega$ results in the following GPE:
\begin{equation}
\begin{split}
    \eye\frac{\pa \psi(x)}{\pa t} =&\Bigg[-\frac{1}{2}\frac{\pa^2}{\pa x^2}+\eye \Gamma\frac{\pa}{\pa x} +\frac{q}{\sigma \sqrt{2\pi}}e^{-x^2/2\sigma^2} \\
  &+(\nb-1)\frac{q}{\sigma \sqrt{2\pi}}e^{-(x\pm L/2)^2/2\sigma^2} -\left|\psi(x)\right|^2\Bigg]\psi(x),
\end{split}
\label{eqn:SAGGPE}
\end{equation}
where $\Gamma = \Omega L/2\pi$  [which we can also write in terms of the dimensional circumference $L_{\mathrm{D}}$ and angular frequency $\Omega_{\mathrm{D}}$ as $\Gamma = (\hbar/|\gd |N)\Omega_{\mathrm{D}}L_{\mathrm{D}}/2\pi$], and $\psi$ is  the (unit norm) condensate wave function.  Note the two barrier terms; presence or absence of the second barrier implies two different forms of Sagnac interferometry: one where both solitons perform full circumnavigations of the ring, enclosing the area within the ring twice; and one where each soliton circumnavigates a different half of the ring, enclosing the area once. These cases are distinguished by $\nb=1$ for the first (single barrier) case and $\nb=2$ for the second (two antipodal barriers) case; the second barrier term is zero for $\nb=1$ [see~\figreft{fig:overview}{(a)}] and identical to the other barrier term, up to a spatial offset, for $\nb=2$ [see~\figreft{fig:overview}{(b)}]. All simulations were carried out with $\sigma=0.2$; this width is suitably narrow to approximate a delta function for collisional velocities up to $v=4.0$~\cite{Note2}, corresponding to a (variable, depending on the ring circumference) dimensionless angular velocity of $\omega=2\pi v/L$.

\begin{figure}[t]
    \centering
    \includegraphics[width=\columnwidth]{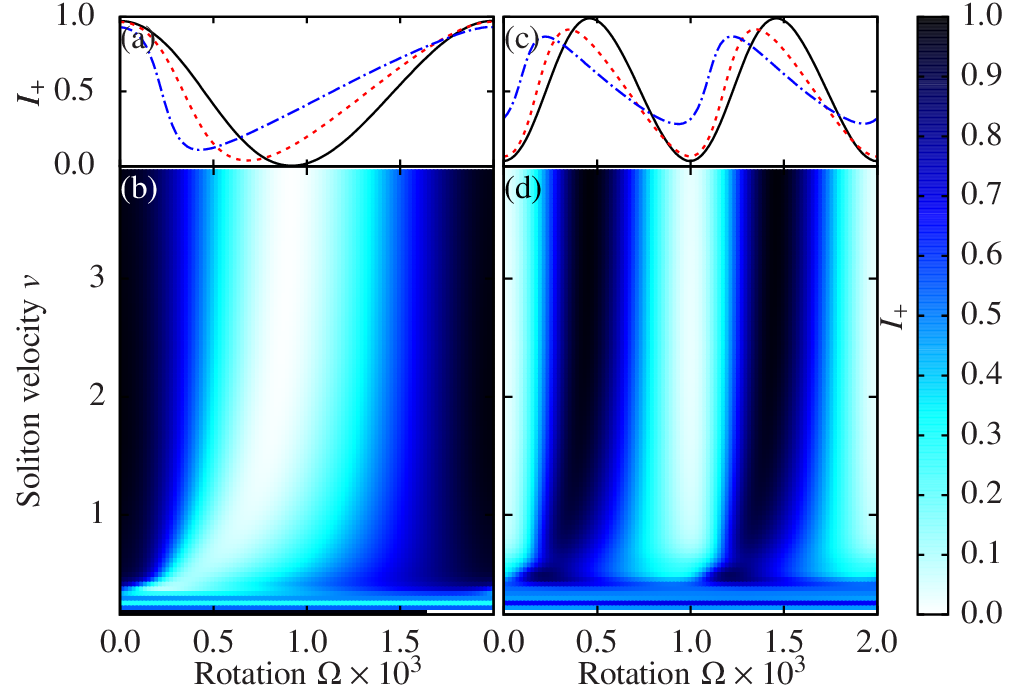}
    \caption[Sagnac interferometry: numerical treatment]{Numerically calculated transmission into the positive domain after the second collision, $I_+$, for the two Sagnac interferometry geometries shown in Fig.~\ref{fig:overview}. Colormaps for the (b) two barrier and (d) single barrier cases show the $0.16<v<4$, $0<\Omega\times10^3<2.5$ parameter space. Panels (a) and (c) show specific curves of constant $v$ for these scenarios [for $v=0.52$ (dashed-dotted line), $v=1$ (dashed line), and $v=4$ (solid line)], and highlight how the different interrogation times result in a different Sagnac phase accumulation. The phase difference is varied by varying $\Omega$ while keeping $L=40\pi$ [hence the $v$ ranged over in panels (b) and (d) correspond to dimensionless angular velocities of between $\omega = 0.008$ and $\omega = 0.2$].
}
\label{fig:int_s}
\end{figure}

We obtain soliton solutions to~\eqnreft{eqn:SAGGPE} (in the absence of splitting potentials and periodic boundary conditions), i.e., the usual nonlinear Schr\"{o}dinger equation in a frame moving with velocity $\Gamma$, by the Galilean invariance of the standard soliton profile~\cite{gordon_ol_1983}. The (amplitude $A$) invariant soliton solution is $\tilde\psi(\tilde x,t)=A\sech(A[\tilde x-Vt])\exp(\eye V\tilde x + \eye[A^2-V^2]t/2)$; the tilde notation denotes the \textit{stationary\/} frame of reference. A soliton moving with velocity $v$ in a frame moving with velocity $\Gamma$ is moving at velocity $V=v+\Gamma$ in the stationary frame. In the \textit{moving\/} frame, where $x=\tilde x-\Gamma t$, we obtain
\begin{equation}
\begin{split}
\psi(x,t)=&A\sech(A[x-vt])\\
&\times\exp(\eye[v+\Gamma][x+\Gamma t]+\eye\{A^2-[v+\Gamma]^2\}t/2).
\end{split}
\label{eqn:tw_sol}
\end{equation}
Assuming $L\gg1$ (a parameter regime far from the critical point described in \cite{kanamoto_etal_pra_2003}), \eqnreft{eqn:tw_sol} is a valid solution to~\eqnreft{eqn:SAGGPE}.
  
\begin{figure}[t!]
  \centering
  \includegraphics[width=\columnwidth]{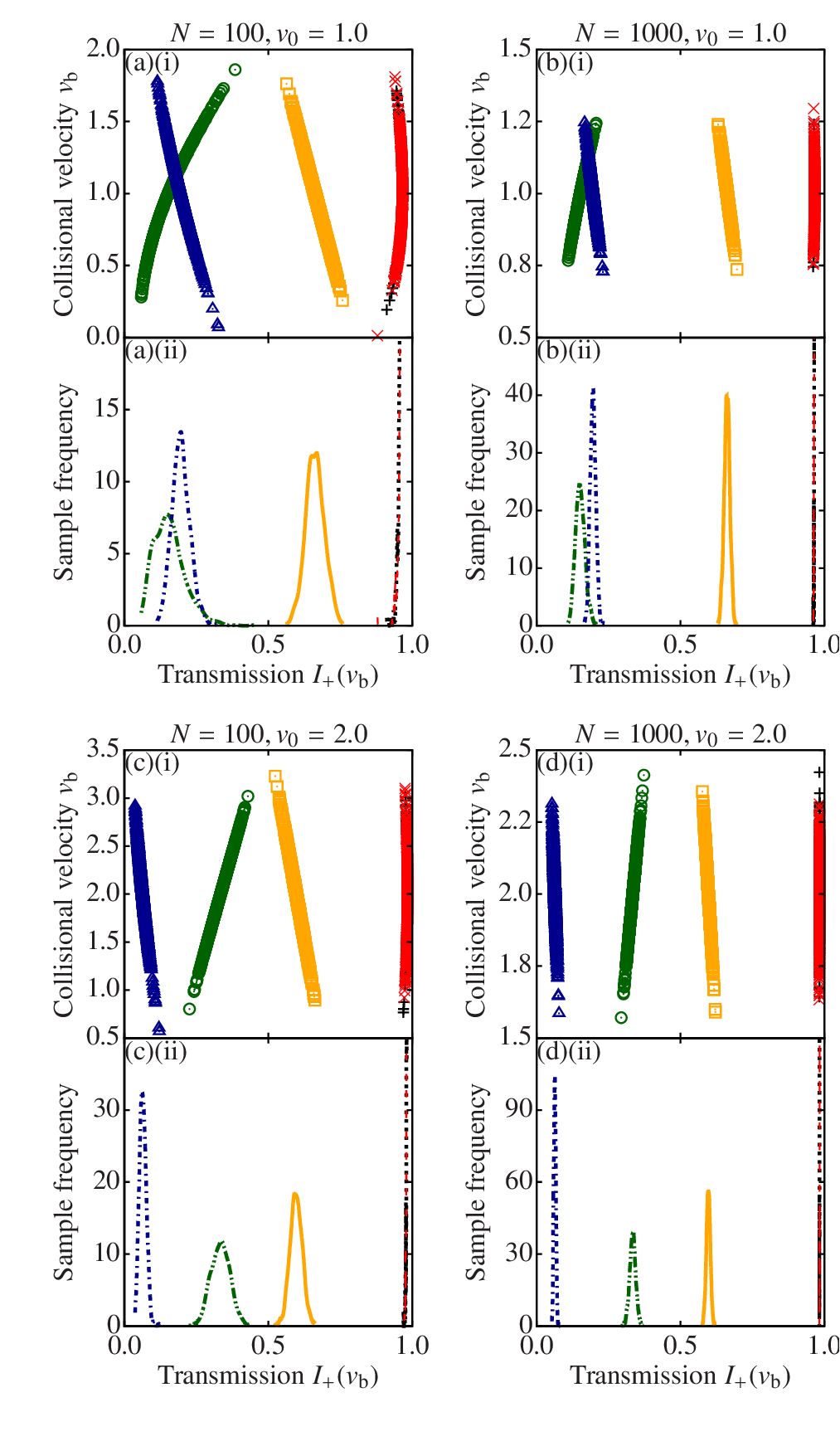}
  \caption[Monte Carlo analysis]{Results of Monte Carlo simulations used to model effects of quantum uncertainty for a range of $v_0$, $N$ and $\Omega$. (a--d)(i) Scatter plot of the solitons' collisional velocity $v_{\mathrm b}$ for ensembles of individual simulations. In (d)(i), the higher gradient of the curves through the points implies the detected transmission $I_{+}$ is less sensitive to quantum fluctuations. (a--d)(ii) Sample distributions of the simulation outcomes. For each $N,v_0$ pair we explored $\Omega\times10^3={0, 6.25, 12.50, 18.75, 25}$, corresponding to Sagnac phases of $\Sdelta={0\mcsa,\pi/2\mcsb,\pi\mcsc,3\pi/2\mcsd,2\pi\mcse}$ respectively.  The simulations were carried out in a two barrier system, with $L=40\pi$ (hence the $v_{0}$ values correspond to dimensionless angular velocities $\omega = 0.05$, $0.1$). The $I_{+}$ peak locations are consistent with the GPE-predicted nonlinear skew for these velocities \cite{HELM_PRA_2012} (see also Fig.~\ref{fig:int_s}).}
\label{fig:mc} 
\end{figure}

We now outline the three-step process of soliton Sagnac-interferometry, common to both ($\nb=1,2$) configurations; later we will analyse the system phase evolution fully.  First, a ground state soliton is split into two secondary solitons, of equal size and a specific relative phase, at a narrow potential barrier [time $\Ts$ in~\figreft{fig:overview}{(a)(ii) and (b)(ii)}]. We obtain an equal split by selecting the barrier's strength $q$~\cite{Note3} for a given incident velocity $v$~\cite{loiko_etal_epjd_2014} and barrier width $\sigma$.  It is the velocity $v$ in the frame \textit{comoving\/} with the barrier that must be known; the value of the frame velocity $\Gamma$ (itself due to the angular frequency $\Omega$) does not affect the outcome. In the second step the secondary solitons accumulate a further relative phase difference $\Sdelta$. This is the $\Omega$-dependent quantity we wish to measure, gained as a result of the differing path lengths travelled by counter propagating waves in a moving frame [time $\Ts<t<\Tc$ in~\figreft{fig:overview}{(a)(ii) and (b)(ii)}]. Finally, the two solitons collide at a narrow barrier [time $\Tc$ in~\figreft{fig:overview}{(a)(ii) and (b)(ii)}]. After this collision the wave-function integrals on either side of the barrier,
$I_{\pm}=\pm\int_0^{\pm L/2}|\psi(x)|^2 \mathop{dx}$,
allow us to determine the value of $\Sdelta$~\cite{HELM_PRA_2012,HELM_PRA_2014}, where $I_+$ and $I_-$ are the positive and negative domain populations. These are ideally determined with an atom number variance below one particle, i.e., exact particle counting at output. This is a challenge facing the whole field of atom interferometry, particularly for experiments pursuing Heisenberg-limited measurements. Single atom resolution has been achieved using a variety of techniques \cite{ockeloen_etal_pra_2010,muessel_etal_apb_2013,bucker_etal_njp_2009,hu_etal_ol_1994,alt_etal_pra_2003,serwane_etal_science_2011,grunzweig_etal_np_2010,puppe_etal_prl_2007,gehr_etal_prl_2010,goldwin_etal_nc_2011} for small numbers ($N\sim 10$) and has recently \cite{hume_etal_prl_2013} been extended to mesoscopic ensembles ($N\sim 1000$) typical of the output states of the soliton interferometer.

To determine how the Sagnac effect manifests in GPE soliton interferometry, we must describe the phase dynamics more fully.   After the initial split at time $\Ts$, the transmitted soliton (in the positive domain) has peak phase $\pT(t)$ (value of the phase at the position of the soliton's peak amplitude), while that reflected (in the negative domain) has peak phase $\pR(t)$. We wish to determine the phase difference $\Delta$ between the solitons before they collide 
with one another at a barrier at time $\Tc$, i.e., $\Delta=(-1)^{\nb}[\pT(\Tc) - \pR(\Tc)]$ [the prefactor $(-1)^{\nb}$ changes the sign of the phase difference to account for the solitons approaching the collisional barrier from different directions depending on the value of $\nb$].  In both cases we choose $\Ts=L/4v$ (the initial soliton starts at $x=-L/4$). If $\nb=1$ the solitons created by the splitting event must both fully circumnavigate the ring before colliding at the same barrier, while for $\nb=2$ the solitons only travel half the distance; hence $\Tc=\Ts+L/\nb v$. In the limiting case of a $\delta$-function barrier, the first (splitting) step causes the transmitted soliton to be phase shifted by $\pi/2$ ahead of the (equal amplitude) reflected soliton, as shown analytically in~\cite{HELM_PRA_2014}. We use this figure as an estimate of the phase difference accumulated by splitting on a Gaussian barrier~\cite{HELM_PRA_2012}; see~\cite{POLO_etal_PRA_2013} for a discussion of phase shifts accumulated with finite-width barriers. We select a Gaussian profile for the barrier, as is typical for experimental setups involving off-resonant sheets of light~\cite{Marchant_etal_2013}, and take $\pT(\Ts)=\pR(\Ts)+\pi/2$. We obtain the phase evolution at the peak of an individual soliton by taking the imaginary part of the exponent of~\eqnreft{eqn:tw_sol} and setting $x=vt$, giving (up to an initial offset) $\pS(t;v)=[A^2+(\Gamma+v)^2]t/2$. Hence, $\pR(t)=\pS(t-\Ts;-v)$, $\pT(t)=\pS(t-\Ts;v)+\pi/2$, and the final phase difference between the solitons is 
\begin{equation}
\Delta=(-1)^{\nb}(2\Gamma L/\nb+\pi/2).
\label{eqn:delta_fin}
\end{equation} 

Without a second barrier ($\nb=1$), the solitons mutually collide at the point antipodal to the splitting barrier.  As this occurs in the absence of any axial potentials or barriers, the solitons are unaffected beyond asymptotic shifts to position and phase~\cite{gordon_ol_1983,zakharov_shabat_1972}, given by
\begin{equation}
   A_j\delta x_j+\eye\delta\phi_j=(-1)^k\ln\left(\frac{A_j+A_k+\eye[v_j-v_k]}{A_j-A_k+\eye[v_j-v_k]}\right)
\label{eqn:asymptotic_2}
\end{equation}
where $j,k \in \{1,2\}$ and $j\ne k$. The quantities $\delta x_j$ and $ \delta\phi_j$ are asymptotic position and phase shifts associated with the $j$th soliton, while $v_j$ and $A_j$ describe that soliton's velocity and amplitude. Associating the soliton transmitted through the barrier at time $\Ts$ with $j=1$, we obtain the correct sign for our asymptotic shifts.  In our case, noting that $A_1=A_2=1/4$ we determine the relative phase shift, and the relationship between the position shifts which arise as a result of this collision to be:
\begin{equation}
\begin{split}
  &\pC=\delta \phi_2 - \delta \phi_1
     =
     \Im\{\ln(16v^2/[16v^2+1])\}
    =0,
\\
  &A_j\delta x_j =
  [(-1)^{k}/2]\ln(1+1/16v^2)
                =-A_k\delta x_k.
\end{split}
\label{eqn:pass}
\end{equation}
Both results use the standard complex logarithmic identity $\ln(z)=\ln(|z|^2)/2+\eye\arg(z)$. \eqnreftfull{eqn:pass} shows us that $\pC$ can be omitted from the calculation of $\Delta$, that $\delta x_j\to 0$ rapidly as $v\to \infty$, and also that whatever the size of the asymptotic position shift, the solitons are always shifted by equal amounts in opposite directions, and so will always meet at the collisional barrier situated at $x=0$. Hence, the antipodal collision in the absence of a barrier does not affect the outcome of Sagnac interferometry if we can assume that the solitons' accelerations during the collision do not affect the Sagnac phase accumulation. The analysis supporting this assumption is beyond the scope of the current work but can be verified numerically.  A potential experimental advantage of the single-barrier configuration is that there is no need to locate a second barrier with great precision relative to the first; that both splitting products traverse exactly the same path before recombining is also likely to ``smooth over'' effects of small asymmetries in the trapping potential.

We can now determine $I_\pm$ by recalling previous results pertaining to soliton collisions at narrow barriers~\cite{HELM_PRA_2012}. Following the same procedure outlined in \cite{HELM_PRA_2014} we obtain 
  \begin{equation}
    I_\pm =[1 \pm (-1)^{\nb} \cos(\Sdelta-\epsilon)]/2,
    \label{eqn:pop_pm}
  \end{equation}
where $\epsilon\to0$ as $v\to\infty$, and 
$
\Sdelta=|\Delta| - \pi/2 = 2\Gamma L/\nb
       =\Omega L^2/\pi\nb
       =(m/\hbar)(\Omega_{\mathrm D} L_{\mathrm D}^2/\pi\nb)
$ is the Sagnac phase we wish to determine. We show results of numerical GPE simulations in~\figreft{fig:int_s}{(a,b)}. For very high velocities, $v\approx4$, the interference follows our prediction~\eqnrefp{eqn:pop_pm} closely, with very small skews arising from nonlinear effects during the final barrier collision, i.e., we can consider $\epsilon\approx0$ in this regime. The $\nb=1$ (c,d)  and $\nb=2$ (a,b) cases have similar structures, however for $\nb=1$ the phase varies twice as quickly, as the interrogation time per shot is twice as long. Otherwise, the similarity of the structures supports the assumption that accelerations during barrier-free collisions do not affect the Sagnac phase accumulation.  As we reduce the velocity, and the necessary (to avoid complicating nonlinear effects arising from a slow interaction with the barrier) assumption of high initial kinetic energy \cite{holmer_etal_cmp_2007,holmer_etal_jns_2007} breaks down, our numerics show that the preceding analysis no longer holds, and so we conclude that Sagnac interferometry is not practicable in the $v\lesssim1$ regime. This is consistent with previous work delimiting $1\geq v\geq 0.25$ as the high-to-low-energy transitional regime~\cite{HELM_PRA_2014}, and the results shown here are comparable to those obtained for the Mach--Zehnder configuration~\cite{HELM_PRA_2014}.  

Figure~\ref{fig:mc} shows results of Monte Carlo simulations following the methodology described in~\cite{HELM_PRA_2014}, which accounts for quantum uncertainty in the initial soliton's center of mass (CoM) position and velocity by adding Gaussian random offsets to the classical soliton's initial velocity and peak position. Here we consider a two-barrier system where the soliton is initially accelerated by a harmonic trap, with frequency $\omega_x$ and its minimum at $x=-L/4$. The soliton is prepared and released from a position $x=-L/4-x_0$ (before quantum fluctuations in the CoM are considered). This harmonic trap is then switched off once the soliton reaches $x=-L/4$, and its velocity is $v_0=\omega_x x_0$. The CoM position and velocity uncertainties contribute to velocity uncertainty at the point of collision, giving collisional velocities $v_{\mathrm{b}}$ that follow a Rician distribution~\cite{HELM_PRA_2014}.  Increasing $N$ reduces the widths of the outcome distributions by reducing the relative significance of quantum fluctuations, hence making the transmission curves~\figrefp{fig:mc}{(a-d)(i)} steeper. As the gradients of these curves asymptote upward, the distributions of the simulation outcomes~\figrefp{fig:mc}{(a-d)(ii)} become narrower. The distributions for the $\Sdelta=\pi/2$ and $3\pi/2$ sets of simulations should, ideally, be centered on $I_{+}=0.5$; these distributions do not have the same location, but approach the ideal ($I_{+}=0.5$) with increasing $v_0$. This is due to the nonlinear skew interfering with the phase evolution during the final collision at time $T_{\mathrm{c}}$, as described in~\cite{HELM_PRA_2012}, and predicted by the GPE.

In conclusion, we have employed a GPE treatment to show how, using a moving bright matter-wave soliton as the initial condition, a matter-wave Sagnac interferometer can be realized within a quasi-1D toroidal trapping configuration (ring trap), in combination with one or two narrow Gaussian barriers due to off-resonant sheets of light.  Although both configurations are in principle equally effective, we note that the single-barrier case is likely less susceptible to systematics due to small asymmetries in an experimental configuration.  We have also explored the effects of quantum fluctuations in the atomic matter-wave's center-of-mass position and velocity; we find that, so long as the initial soliton velocity is sufficiently fast, particle numbers of $N\gtrsim 1000$ suffice to give sharp transmission responses, which can then be interpreted to deduce a Sagnac phase. 

\begin{acknowledgments}
We thank D. I. H. Holdaway, A. L. Marchant, and  C. Weiss for useful discussions, and the UK EPSRC (grant no.\ EP/K030558/1) for support.
\end{acknowledgments}


%

\end{document}